\documentstyle[aas2pp4]{article}
\lefthead{Strickman et al.}
\righthead{2CG 135+1}

\newcommand{\ltsima} {$\; \buildrel < \over \sim \;$}
\newcommand{\simlt}  {\lower.5ex\hbox{\ltsima}}            
\newcommand{\gtsima} {$\; \buildrel > \over \sim \;$}
\newcommand{\simgt}  {\lower.5ex\hbox{\gtsima}}            
\newcommand{\ggg}  {$\gamma$}
\begin{document}
\title{A Multiwavelength Investigation of the
Relationship Between 2CG 135+1 and LSI+61$^{\circ}$ 303}
\author{M.~S.~Strickman\altaffilmark{1},
M. Tavani\altaffilmark{2},
M.~J.~Coe\altaffilmark{3},
I.~A.~Steele\altaffilmark{4},
J.~Fabregat\altaffilmark{5},
J. Mart\'{\i}\altaffilmark{6}, 
J.~M.~Paredes\altaffilmark{7},
P.~S.~Ray\altaffilmark{1,8}}
\altaffiltext{1}{Naval Research Laboratory, Washington, DC 20375-5352;
strickman@osse.nrl.navy.mil}
\altaffiltext{2}{Columbia Astrophysics Laboratory, Columbia University, New
York, NY 10027; tavani@astro.columbia.edu}
\altaffiltext{3}{Department of Physics and Astronomy, University of Southampton,
Southampton, SO17 1BJ, U.K.; mjc@astro.soton.ac.uk} 
\altaffiltext{4}{Astrophysics Group, Liverpool John Moores University,
Liverpool, L3 3AF, U.K.; ias@staru1.livjm.ac.uk}
\altaffiltext{5}{Departmento de Astronomia, Universidad de Valencia, 46100
Burjassot, Spain; fabregat@evalvx.ific.uv.es}
\altaffiltext{6}{CEA/DSM/DAPNIA/Service d'Astrophysique,
Centre d'\'Etudes de Saclay,91191 Gif-Sur-Yvette CEDEX,FRANCE}
\altaffiltext{7}{Departament d'Astronomia i Meteorologia, Universitat de
Barcelona, Av. Diagonal 647, E-08028, Barcelona, Spain; josepmp@mizar.am.ub.es}
\altaffiltext{8}{NRC Research Associate; paulr@rira.nrl.navy.mil}

\begin{abstract}

We present the results of a multiwavelength monitoring campaign
targeting the $\gamma$-ray source 2CG 135+1 
in an attempt to
confirm the association of this object with the radio/Be/X-ray 
binary system LSI +61$^\circ$303.
The campaign included 
simultaneous 
radio, optical, infrared, and hard
x-ray/$\gamma$-ray observations 
carried out 
with a variety of instruments, covering
(not continously)
almost three  binary cycles of LSI +61$^\circ$303 
during the period April-July 1994.
Three separate  OSSE observations of the $\gamma$-ray source were 
carried out, covering different phases of the radio lightcurve.
Hard X-ray/\ggg-ray  emission was detected from the direction of
2CG 135+1 during the first  of these  OSSE observations.
The signal to noise ratio of the OSSE 
observations  was insufficient to 
establish a spectral or intensity correlation of the high-energy emission
 with simultaneous radio, optical and infrared emission
of LSI +61$^\circ$303.
We briefly discuss the theoretical implications of our observations.

\end{abstract}


\section{Introduction}


The $\gamma$-ray source 2CG~135+1 is one of the most prominent 
unidentified $\gamma$-ray sources 
near the Galactic plane. 
Since its discovery by the COS-B satellite 
(Hermsen et al. \markcite{herms77}
1977, Swanenburg et
al. \markcite{swane81}1981), 
no satisfactory explanation has been found for the nature of 2CG
135+1 and its $\gamma$-ray emission mechanism.  The COS-B error box of 
2CG~135+1 contains two 
interesting sources, the radio flaring Be star LSI +61$^\circ$303, 
and the QSO 0241+622 (e.g, Perotti et~al. 1980).
EGRET observations of 2CG~135+1 during CGRO Phase 1 and 2 (von Montigny et al. \markcite{vonmo93}1993, Thompson et al.
\markcite{thomp95}1995, Kniffen et al. \markcite{kniff97} 1997) resulted in
a $\sim 20'$ wide error box location.  The source appeared in the 2nd EGRET
catalog as 2EG~J0241+6119 (Thompson et~al. 1995).
The new EGRET position of the $\gamma$-ray source is well within the
COS-B error box and is consistent with the position of LSI~+61$^\circ$303. 
The position of the QSO 0241+622 is outside the newly established \ggg-ray
error box, being about $50'$ away from the centroid
of the  EGRET error box (Kniffen et~al. \markcite{kniff97}1997).
COMPTEL Phase 1 and 2 observations  also 
confirmed the presence of a \ggg-ray source consistent with the
position of 2EG~J0241+6119 in the
energy range 1--30 MeV (van Dijk et al. \markcite{vandi96}1996).
Due to their limited  angular resolution, COMPTEL observations
 cannot rule out either
LSI +61$^\circ$303 or QSO 0241+622 as a possible counterpart of
the \ggg-ray source.

The remarkable system  LSI +61$^\circ$303 (GT~0236~+610, V615~Cas)
 is one of the few Be/X-ray binary stars
 which exhibits strong radio emission.
The periodic nature of this radio emission was first determined by Taylor \&
Gregory \markcite{taylo82}(1982) who found it  to be strongly modulated with a
period of $\sim$ 26.5 days.  The currently accepted value of the radio period
is $26.496\pm0.008$ days (Taylor \& Gregory \markcite{taylo84}1984).   Radio
phase zero has been arbitrarily set at Julian Date 2443366.775 for the system
with the radio outbursts occuring normally at phases 0.5 -- 0.7.
Spectroscopic radial velocities (Hutchings \& Crampton \markcite{hutch81}1981) 
for the system indicate that the $\sim26.5$ day modulation corresponds to the
orbital period of the binary system.  Modulation of the IR and optical system
at the radio period has been reported by Paredes et al.
\markcite{pared94}(1994).

Optical spectra of the source are
typical of a Be system, showing strong emission in the H$\alpha$ and H$\beta$
lines.  Also typical of a Be star is the infrared excess observed from the
system (Mart\'{\i} \& Paredes \markcite{marti95}1995).  Both of these
observations are generally explained in Be systems by some form of
equatorially enhanced stellar wind (i.e. circumstellar disk) surrounding the Be
star. This circumstellar disk is thought to be the source of the material
accreting onto the neutron star in Be/X-ray binaries.

Soft X-ray emission in the range 0.5-5 keV from LSI +61$^\circ$303 was
detected by {\em Einstein} (Share et al. \markcite{share79}1979, Bignami et
al. \markcite{bigna81}1981), and recent ASCA (Leahy, Harrison \& Yoshida
\markcite{leahy96}1996; Harrison, Leahy \& Waltman \markcite{harri96}1996) and
ROSAT (Goldoni \& Mereghetti \markcite{goldo95}1995, Taylor et~al. 1996)
observations confirm the detection.  Recent analysis of public ASM/RXTE data
(Paredes et al. \markcite{pared97a}1997) has shown that the X-ray emission
also appears to display outbursts with the same period as the radio. These
data show clear evidence for a $26.7\pm0.2$ d period, the folded X-ray light
curve exhibiting a broad high state overlapping the usually active radio
phases.

%
Assuming 2CG~135+1 is associated
with LSI +61$^\circ$303 rather than the QSO, various possible models of the
$\gamma$-ray production mechanism have been proposed.  Recent models include
a non-accreting young pulsar in a system with a mass-losing B-star
(Maraschi \& Treves (1981),
Tavani (1995, 1997)),
and super-Eddington accretion onto a neutron star (Mart\'{\i} \&
Paredes \markcite{marti95} 1995).

In an attempt to confirm the association between the $\gamma$-ray source and
LSI +61$^\circ$303 and to provide input data for modeling of the system,
simultaneous $\gamma$-ray, radio, optical, and infrared  observations were
obtained during the interval 1994 April--July. The OSSE instrument on the {\it
Compton Gamma-Ray Observatory} (CGRO) performed a series of hard
X-ray/$\gamma$-ray  
pointings centered at 2CG~135+1 during
April and again during June and July, 1994. Frequent radio observations of
LSI~+61$^\circ$303  were carried out at the Green Bank
Interferometer (GBI) and during June and July at the VLA. Likewise, optical
spectroscopy was obtained from the Jacobus Kapteyn Telescope (JKT), La Palma
and infrared photometry from the Carlos S\'anchez Telescope (TCS), Tenerife
during late June and early July, 1994. This paper presents the results of all
of these observations and discusses their implications for the various
proposed models of the system.

\section{Observations}

\subsection{OSSE Observations}

\begin{figure}
\plotone{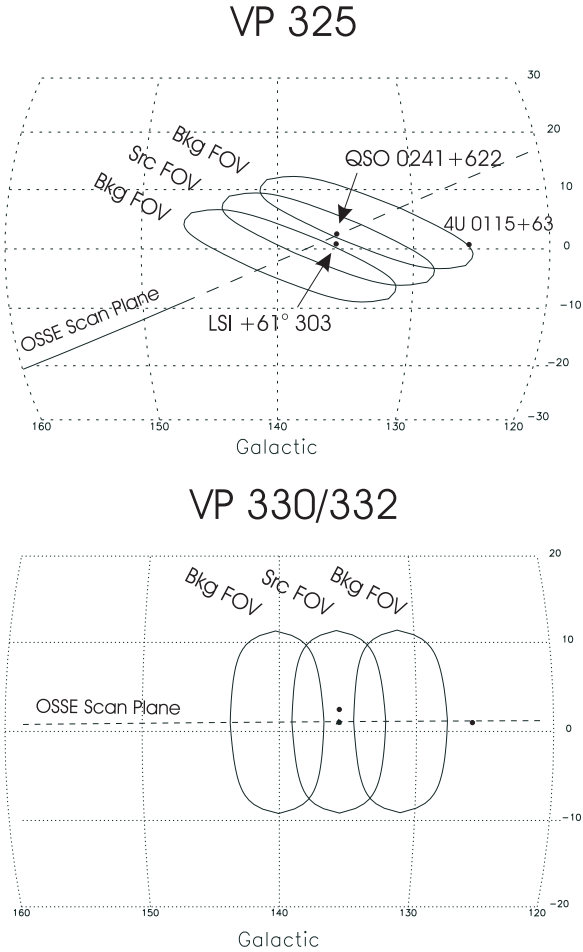}
\caption{\label{fig:osseview} OSSE viewing configuration for VP 325, 330 and
332.  The ovals represent 10\% response contours for an OSSE detector.  In
each case, the source-pointing and two background-pointing fields are
indicated.}
\end{figure}

OSSE observed the region containing LSI +61$^\circ$303 on three
occasions from 1994 April to July, as shown in Table~\ref{tbl:osselog}. 
OSSE, described by Johnson et al. (\markcite{johns93}1993)
consists of four nearly identical NaI detectors, of which two were in use
for these observations.  The OSSE field-of-view (FOV) is $3.^{\circ}8\times
11.^{\circ}4$, large enough that source confusion is a concern, especially near
the galactic plane.  Figure~\ref{fig:osseview} shows the OSSE viewing
configuration for the three observations.  Note that we cannot unambiguously
determine whether any detected emission belongs to LSI +61$^\circ$303 or to
QSO 0241+622 from positional considerations alone.

\begin{figure}
\plotone{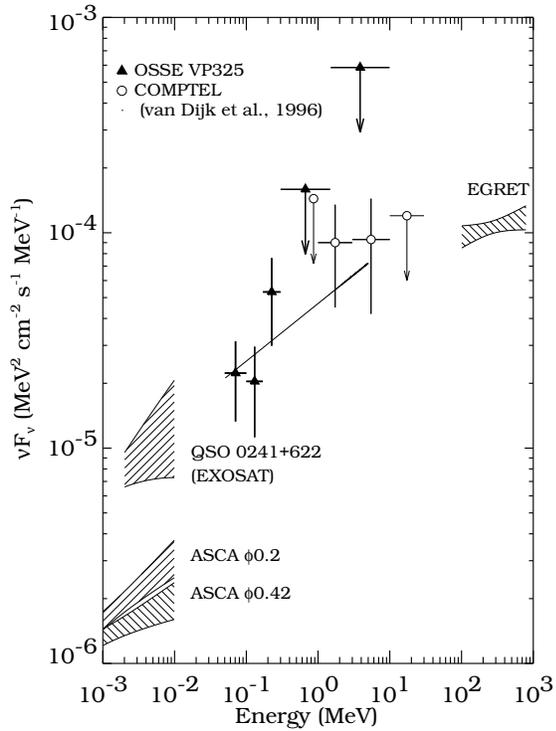}
\caption{\label{fig:ossespec}X-ray and $\gamma$-ray spectra from
the region around
LSI +61$^\circ$303.  Included are data from OSSE VP 325 (triangles together
with best-fit power law) and 
COMPTEL (open circles). Also shown are results on LSI +61$^\circ$303 from
EGRET and ASCA. The lines labeled ``EGRET'' represent a best fit power law 
68\% confidence
interval from Thompson et al. (1995), while the two sets of
lines labeled ``ASCA $\phi0.2$'' and ``ASCA $\phi0.42$'' are 90\% confidence
intervals for ASCA measurements at radio binary phases 0.2 and
0.42 respectively (Leahy et al. 1996).  The lines labeled
``QSO 0241+622'' are
a 90\%  confidence interval of the EXOSAT spectrum
from QSO 0241+622 (Turner \& Pounds 1989).}
\end{figure}

In Figure~\ref{fig:ossegbilc} we present the derived OSSE light curves for the
three viewing periods, together with radio light curves from the GBI.
Figure~\ref{fig:ossespec} displays the OSSE spectrum from VP 325.   OSSE
detected a signal with $\simeq 4\sigma$ significance from the direction of
LSI +61$^\circ$303 in the 50--300 keV band during VP 325.  No significant signal
was detected from this region during VP 330 or VP 332.  The resulting upper
limits, listed in Table~\ref{tbl:osseflux}, are consistent with the VP 325
detection, hence the OSSE data do not strongly support a claim of time
variability from one viewing period to the next, nor is there any obvious
correlation between the hard X-ray and radio light curves.  However, we also
note that the OSSE data for VP 325 are inconsistent with a constant flux at
the 99\% confidence level. 

The OSSE spectrum, shown in Figure~\ref{fig:ossespec}, is well represented by
a power-law model with photon index 1.6$\pm$0.6 over the 50--300 keV range. 
In order to estimate the contribution of the diffuse galactic continuum
emission to the observed signal, we have used a  model normalized to OSSE
galactic plane observations (Skibo 1996, private communication).  The
resulting estimated contribution, at least an order of magnitude less than the
observed flux, is negligible.  In addition, contributions from the binary
X-ray pulsar 4U 0115+63, which was located on the edge of one of the
background fields during VP 325 (see Figure~\ref{fig:osseview}), are
negligible.  A major outburst from this source took place just after the end
of VP 325
(Negueruela, et al. \markcite{negue97}1997).  Although the source was still
active during VP 330 and VP 332, the viewing configuration for these periods
excluded 4U0115+63 as a potential contaminant. 

\subsection{Radio Observations}
\subsubsection{Green Bank Interferometer}
The Green Bank Interferometer (GBI) began a program of continuous
daily monitoring of LSI +61$^\circ$303 in 1994 January which continued until 1996
February.  For a complete description of these observations see Ray
et al. \markcite{ray97}(1997).  Briefly, the GBI consists of two 26 m antennas
on a 2.4 km baseline, each of which has a pair of cooled 35 MHz
bandwidth receivers to simultaneously receive signals at 2.25 and 8.3
GHz.  Ten minute observations of LSI +61$^\circ$303 were performed 1--10 times per
day within $\pm 5$ hours of source transit.  Measured correlator
amplitudes are converted to flux densities by comparison to standard,
regularly observed calibrators.  Flux densities from the GBI
monitoring are available during all of the OSSE viewing periods.
Figure~\ref{fig:ossegbilc} shows daily averages of the GBI data.

\subsubsection{VLA Observations} \label{intro} In addition to the Green Bank
Data, the VLA interferometer was used to monitor the radio emission of LSI
+61$^\circ$303 throughout a full orbital period of 26.5 d concurrent with the
OSSE observations in VP 332. Sixteen VLA observing sessions were carried out,
evenly spaced by a few days, and covering the time interval from 1994 June 9
to 1994 July 8. Each session had a typical duration between 2--4 h and allowed
us to measure the  LSI +61$^\circ$303 flux density at up to 5 different radio
wavelengths (20, 6, 3.5, 2.0 and 1.3 cm).  The VLA was always in the B
configuration.  During the two shortest sessions available only three
wavelengths were observed (20, 6 and 3.5 cm), while in some cases it was
possible to include a few additional 90 cm measurements.  For all
observations, a bandwidth of 100 MHz was employed except at 90 cm, where a
smaller bandwidth of 6 MHz was selected.  The data were edited and calibrated
using the AIPS software package.  The source 3C48 was always used as absolute
amplitude calibrator, while the phase calibrator adopted at all wavelengths
was 0228+673, except at 90 cm where 0229+777 was used instead.  In most
observations LSI +61$^\circ$303 appeared as an isolated strong point source,
allowing us to use phase self calibration to improve the dynamic range of the
maps from which flux densities were measured.

\begin{figure}
\plotone{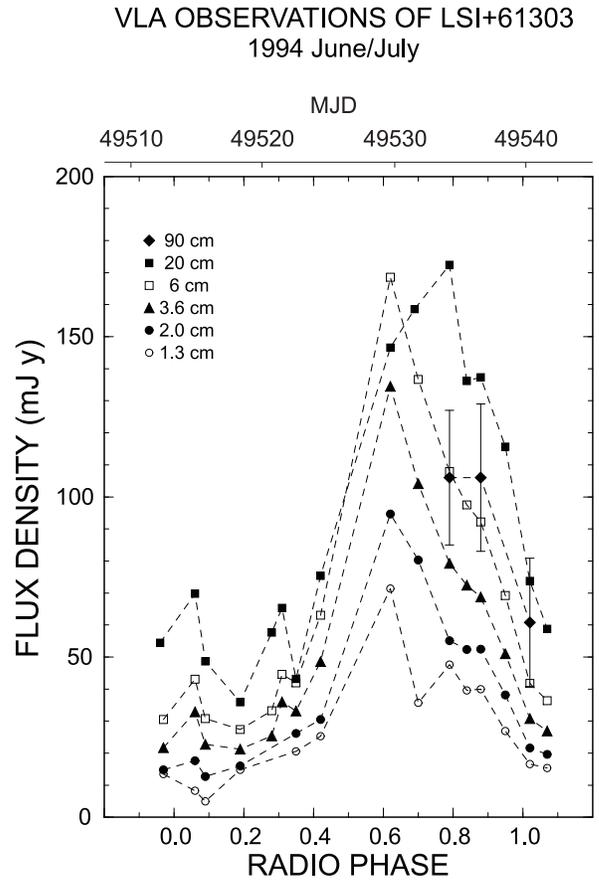}
\caption{\label{fig:curves}
Multifrequency radio light curves of LSI +61$^{\circ}$303 observed
with the VLA corresponding to the radio
outburst of 1994 June-July. Error bars not shown are smaller
than the symbol size.}
\end{figure}

The results of the VLA radio monitoring are given in Table \ref{radiodata}. 
The flux density errors quoted in columns three through eight are the rms
noise of the map in mJy beam$^{-1}$.  The data listed in Table \ref{radiodata}
are also plotted in the panels of Figure~\ref{fig:curves} in the form of radio
light curves for the different wavelengths, showing the development of the
radio outburst.  The time evolution of the radio spectrum is displayed in the
panels of Figure~\ref{fig:specevol} for each observing session. A power law
model spectrum has been fit to each spectrum.  Significant phase-dependent
deviations from the power law, especially at the lowest and highest
frequencies, are evident in this figure.  In order to discuss these effects in
more detail, we have plotted spectral indices between adjacent wavelength
pairs, in addition to the best fit power law index, as a function of radio
phase (Figure~\ref{fig:specind}).

\subsection{Optical Spectroscopy}

\begin{figure}
\plotone{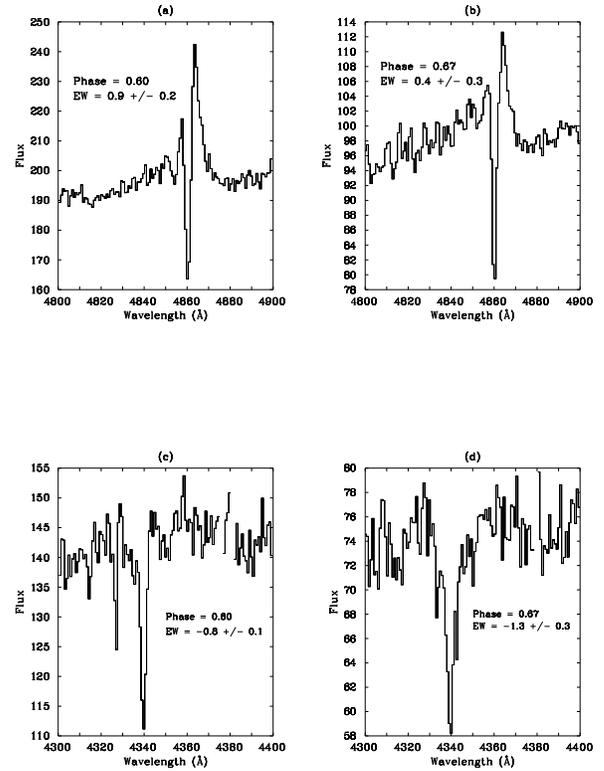}
\caption{\label{fig:hbetaspec}Details of the H$\beta$ (panels (a) and (b))
\& H$\gamma$ (panels (c) and (d)) profiles taken at two different binary
phases.  The vertical scales are in arbitrary flux units.}
\end{figure}

Optical spectra of LSI +61$^\circ$303 in the regions of the H$\alpha$,
H$\beta$ and H$\gamma$ lines were obtained during the period 1994 June 25 --
28.  The observations were made using the Richardson-Brearly Spectrograph
(Edwin \markcite{edwin88}1988)  of the Jacobus Kapteyn Telescope, La Palma
with the EEV7  CCD detector.  The dispersion was 0.8{\AA} per pixel, and, from
observations of narrow interstellar features in our spectra, we estimate the
spectral resolution was less than 2{\AA} in all our observations. A log of the
observations is given in Table 6.  Data reduction was carried out according to
standard spectroscopic procedures  using the STARLINK supported package FIGARO
(Shortridge \markcite{short91} 1991).  The spectra were flux calibrated with
observations of standards from Oke \& Gunn \markcite{okegu83}(1983).  No
attempt was made to remove atmospheric absorption features from the spectra
because these lines do not contaminate the regions of interest for this work.  
In Figures~\ref{fig:halphaspec} and \ref{fig:hbetaspec}, we plot excerpts
from the spectra showing the H$\alpha,\beta$ and $\gamma$ lines.
We discuss model fitting to these spectra in a later section.

\subsection{Infrared Photometry}
\begin{figure}
\plotone{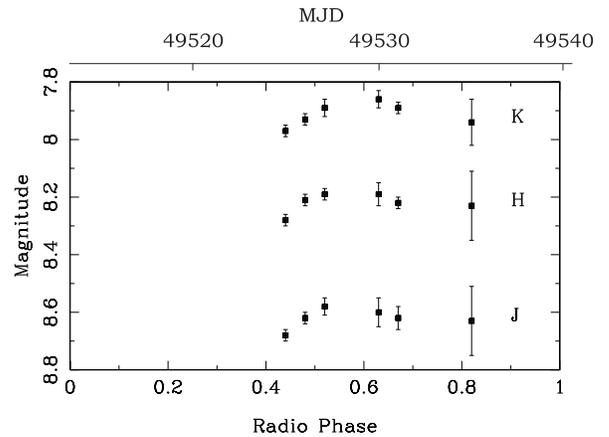}
\caption{\label{fig:irlc}Infrared light curves of LSI +61$^\circ$303}
\end{figure}

Infrared $J$ ($\sim 1.2\mu$m), $H$ ($\sim1.6\mu$m) and $K$ ($\sim
2.2\mu$m) observations of LSI +61$^\circ$303 were obtained from the Carlos
S\'anchez Telescope (TCS), Tenerife during the period 1994 June 21 --
1994 July 1 using the CVF infrared photometer.  The observations
are listed in Table~\ref{irtable}
and plotted in Figure~\ref{fig:irlc}.  Observations on the last night are of
poorer quality due to dust in the atmosphere.  A clear trend in the
data is apparent, with the infrared flux brightening during the period
from radio phase $\sim 0.4$--$0.6$.  At later phases, our data are consistent
with either constant or declining infrared flux.

\section{Discussion}

\subsection{The OSSE Results}

To place the OSSE observation in the context of other observations at
different energies, Figure~\ref{fig:ossespec} includes noncontemporaneous
results from COMPTEL (Van Dijk et al. \markcite{vandi96} 1996), EGRET
(Thompson et al. \markcite{thomp95} 1995), and ASCA (Leahy et al.  
\markcite{leahy96} 1996).   (Note that COMPTEL cannot resolve LSI
+61$^\circ$303 from QSO 0241+622).  Although we have not displayed the MISO
result (Perotti et al. \markcite{perot80}1980),  it clearly exhibits a higher
flux than either the OSSE or COMPTEL observations.  However, due to the low
detection significance of all these data, it is difficult to conclude that the
MISO result is inconsistent at more than a $3\sigma$ confidence level.  We
have extrapolated the ASCA 1--10 keV best fit power law models from two
observations (radio phase 0.2 and 0.42) into the OSSE energy range. 
Observations by Goldoni \& Mereghetti (\markcite{goldo95}1995) using the ROSAT
PSPC cover too low an energy range to usefully extrapolate to $\gamma$-rays,
especially considering possible correlation of column density and power law
index in their best fit result.  The ASCA result, on the other hand, extends
well above the energy band where absorption is effective, and therefore
supplies a better measure of the underlying power law.  As suggested by Leahy
et al. \markcite{leahy96}(1996), the extrapolation predicts considerably less
emission in the OSSE range than was observed.  The OSSE integral flux from
50--300 keV during VP 325 is $(4.3\pm1.1)\times10^{-4}$ photons cm$^{-2}$
s$^{-1}$, while the integral of the ASCA best fit power law models
extrapolated to the same energy range are  $(1.0\pm0.4)\times10^{-4}$ and
$(0.5\pm0.2)\times10^{-4}$  photons cm$^{-2}$ s$^{-1}$ for radio phases 0.2
and 0.42 respectively.  These are inconsistent at the 2.8$\sigma$ and
3.4$\sigma$ levels, assuming that the spectrum does not change (in particular,
does not harden significantly) from 10 keV to 300 keV.

Since the relatively large OSSE field-of-view does not exclude the nearby QSO
0241+622 as a possible source of emission, we have also plotted the result of
EXOSAT (2 -- 10 keV) observations of the QSO (Turner \& Pounds 
\markcite{turne89} 1989), corrected for the its position in the OSSE
collimator response.  Note that the 1.7 photon index power law spectrum used
to model the EXOSAT data is consistent with the the OSSE result.

When we look at an overall view of the spectra from this region, no clear
picture emerges.  However, the fact that OSSE and COMPTEL results fall
significantly above an extrapolation of the ASCA spectra and
that the OSSE and COMPTEL spectra are
consistent with the X-ray spectrum of QSO 0241+622, lead to the likelihood
that at least some, and perhaps a significant amount, of the flux observed by
both OSSE and COMPTEL
might come from the QSO rather than LSI +61$^\circ$303.  Cycle-to-cycle
variations in the X-ray emission may also contribute to the observed
discrepancy.  More contemporaneous X-ray and \ggg-ray data are required to
test for the existence of cycle to cycle variability of the LSI +61$^\circ$303
system and to better study the X-ray period recently reported by Paredes et
al. (\markcite{pared97a}1997).

Similarly, the low significance of the  OSSE light curves 
shown in Figure~\ref{fig:ossegbilc} does not
allow us to use temporal signatures to associate LSI +61$^\circ$303
with the OSSE source.   
We note that the flux detected by OSSE is relatively higher
during the April 1994 observation, in apparent coincidence
 with a prominent onset and slow  decay of a radio flare.
June 1994 observations show a null detection in the BATSE
energy range in coincidence with an extended minimum of the
radio light curve. July 1994 observations by OSSE show a fluctuating
low-intensity  \ggg-ray flux in coincidence with the onset and decay
of another radio flare.  We see little of no evidence of $\gamma$-ray emission
before the radio flare during July, although the X-ray evidence (e.g. Paredes
et al. \markcite{pared97a}1997, Leahy et al. \markcite{leahy96}1996) indicates that the high energy flare preceeds
the radio.
%
We are therefore unable to use either spectral
or temporal OSSE information  to unambiguosly
associate the $\gamma$-ray
emission from this region of the sky with LSI +61$^\circ$303.  

\subsection{Model Constraints}
Assuming 2CG 135+1 is associated with the radio/X-ray
source, we may ask if our observations place any constraints on
two proposed models of the system currently
under debate.  

In the first model LSI~+61$^\circ$303 is assumed to
contain a non-accreting young pulsar in orbit around the mass-losing
Be-star (e.g. Maraschi \& Treves \markcite{maras81}1981, Tavani 
\markcite{tavan95}1995).  In this case the
high energy emission results from the interaction of the pulsar wind
with the circumstellar material surrounding the Be star, a mechanism
which has recently been shown to operate in the Be star/pulsar system
PSR 1259-63 (Tavani \& Arons \markcite{tavan97}1997).  
The modulation of the radio
emission may be due to the the time-variable geometry of a 
`pulsar cavity' as a function of orbital phase.  Ultimately the
high-energy shock emissivity will depend on the geometrical and
radiative characteristics of the pulsar cavity as the pulsar
orbits around the primary.  Long timescale modulations in the emission
from the system are explained by the influence of a time
variable mass outflow from the Be star, and its effect on the
size of the pulsar cavity and its emissions.

The second model for the LSI~+61$^\circ$303 system 
proposes that the high energy emission 
results from super-Eddington
accretion of the circumstellar material surrounding the Be star onto
a neutron star  (e.g. Taylor et al. \markcite{taylo92}1992, 
Mart\'{\i} \& Paredes \markcite{marti95}1995).
As well as producing the high energy photons, this is
assumed to result in a stream of relativistic particles which cool
by synchrotron emission, resulting in the radio emission.  The
modulation of the radio light curve is assumed to be due to the
relativistic electrons only being accelerated during a short period
(or periods) of each orbit, with wind optical depth effects dictating
the time delay between the injection of the electrons and the visibility of
the radio emission.  

We shall now consider (at least qualitatively) whether each of the
proposed models can explain the observed features of the phase
resolved radio and X-ray spectra.

\subsubsection{Non-accreting pulsar model}

The pulsar model for 2CG~135+1 is based on a {\it non-accreting}
high-energy source. A rapidly rotating pulsar, as expected in
young binary systems located near the Galactic plane, can be
sufficiently energetic to avoid accretion of gaseous material
provided by the mass outflow of a massive companion star.
A relativistic pulsar wind, made of electron/positrons (possibly
ions) and electromagnetic fields can sweep away the
gaseous  material from the Be star companion and produce
high-energy emission by a relativistic shock.
This mechanism of high-energy emission is believed to power
the unpulsed X-ray and \ggg-ray emission from the Crab nebula
(Nolan et  al., 1993), and it was recently demonstrated to
fit the time variable high-energy emission
observed near periastron from the Be star/pulsar system
PSR~B1259-63 (Tavani \& Arons 1997, hereafter TA97).
PSR~B1259-63, a 47~ms pulsar of moderately high spindown power
in a long-period (3.4~yrs), high eccentricity ($\sim 0.87$) orbit
around a Be star, is particularly suited for
our discussion. 
Time variable soft and hard X-ray energy emission has been detected
throughout the PSR~B1259-63 orbit, with luminosities
in the range $10^{33}-10^{34} \, \rm erg  \, s^{-1}$ (TA97 and references
therein). All the characteristics of the high-energy emission from
this system are in agreement with the expectations from a
shock-driven emission mechanism.  In particular, the lack of
strong X-ray emission, lack of Fe line emission,
  and the small and constant column density 
of the PSR~B1259-63 system throughout its elliptical orbit 
($N_H \sim 6\cdot 10^{21} \rm \, cm^{-2}$, comparable to
LSI~+61$^\circ$303, see
Leahy et al., \markcite{leahy96}1996) strongly argue against an accretion-driven
interpretation of the observations.
In  the case of the PSR~B1259-63 system, the pulsar and 
binary characteristics  conspire to produce a synchrotron
power-law emission up to $\sim 1$~MeV and negligible inverse
Compton scattering in the EGRET energy range, in agreement with
observations (TA97). However, the features of the high-energy
emission of the PSR~B1259-63 system are not unique, and emission
in the   COMPTEL-EGRET energy range is expected  from a more
energetic pulsar orbiting around a Be star (Tavani 1995, 1997). 
2CG~135+1 may  be such a system.
In analogy with the case of the  PSR~B1259-63 system, the low-energy
emission below $\sim 1$~MeV can be powered by synchrotron emission 
from a shock at intermediate distance between the pulsar and the
Be star surface. Emission in the EGRET energy range can be produced 
by inverse Compton scattering of soft photons (both from the
synchrotron component at the shock and from the Be star surface and/or
equatorial disk) against the relativistic particle flow.
Modulation of the high-energy flux from 2CG~135+1 is expected in
this model, depending on the geometry of the pulsar cavity
(Maraschi \& Treves \markcite{maras81}1981) and/or pitch angle anisotropy distribution
of relativistic particles advected away from the shock site (TA97).
An overall time variability of the high-energy flux is also expected
as a function of a variable mass outflow rate from the Be star companion
(Tavani 1995, 1997).   In this model, radio emission can result
from the relativistic particles of the pulsar wind advected away from
the `nose' of the pulsar cavity. 
This radio emission (a combination of synchrotron and inverse
Compton emission from moderately relativistic electrons/positrons in
the pulsar wind) is expected to be enhanced at
 selected orbital  phases depending on the geometry.
Substantial radio enhancement and   `flaring' can occur in
coincidence with orbital phases corresponding to the line of sight
being almost aligned with the axis of the `comet-like' pulsar cavity
in the Be star outflow. Many of the considerations of the next
section on 
synchrotron and inverse Compton losses of radio emitting particles
can be applied also in this context.

Our OSSE observations constrain the synchrotron contribution
to the emission from 2CG~135+1 to be marginally consistent and
not larger than  what observed in the 50-200~keV   range in the 
case of the periastron passage of PSR~B1259-63.
The hard X-ray emission from 2CG~135+1 in coincidence with the
onset and decay of the radio flares of LSI +61$^\circ$303 
(not necessarily corresponding to the periastron) is
therefore constrained to be below $10^{34} \rm \, erg \, s^{-1}$.
We notice here that a featureless power-law
spectrum in the X-ray range is what is expected by a relativistic
shock model of emission in a pulsar binary (TA97).

\subsubsection{Super-Eddington model}

\begin{figure}
\plotone{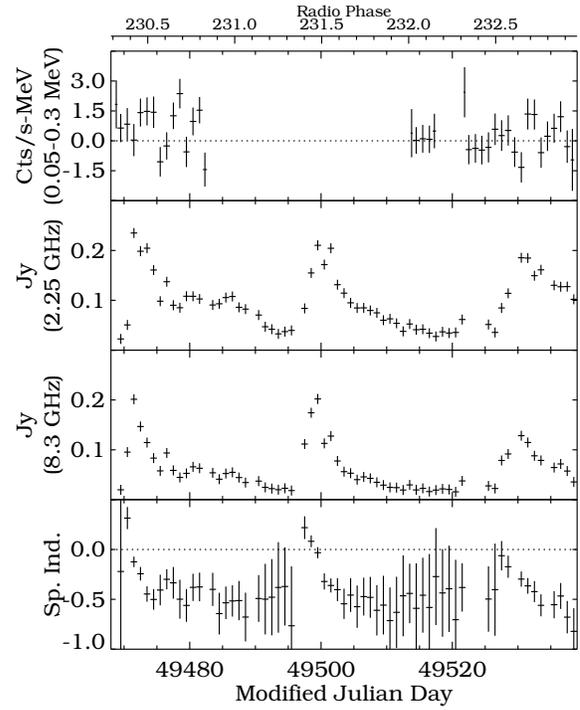}
\caption{\label{fig:ossegbilc}OSSE and GBI lightcurves.  The lefthand group of
OSSE points are VP 325, the first group on the right is VP 330 and the second
group on the right is VP 332.  The abscissa is labeled  both with Modified
Julian Day and with phase of the 26.496-day radio period 
(Taylor \& Gregory 1984). The integer part of
the phase is the number of cycles since the radio ephemeris epoch.}
\end{figure}

\begin{figure}
\plotone{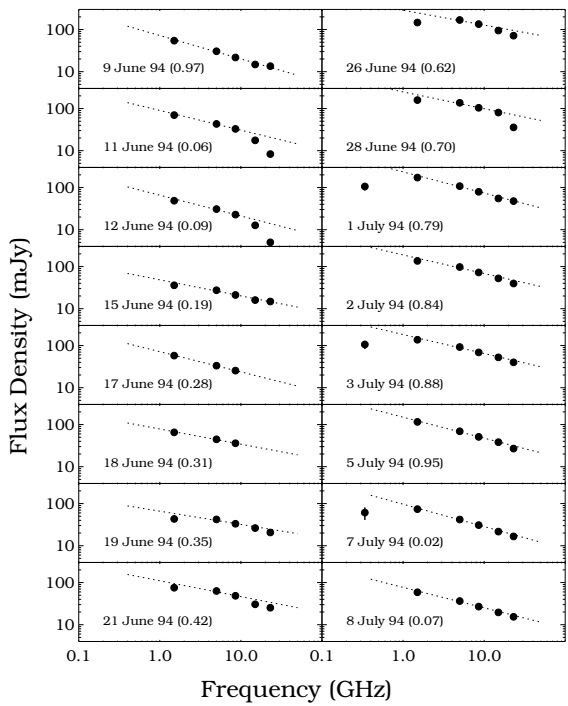}
\caption{\label{fig:specevol}
Time evolution of the LSI +61$^{\circ}$303 non-thermal radio spectrum
during the outburst of 1994 June-July. Spectra are labeled according
to the date and radio phase of the observation.}
\end{figure}

\begin{figure}
\plotone{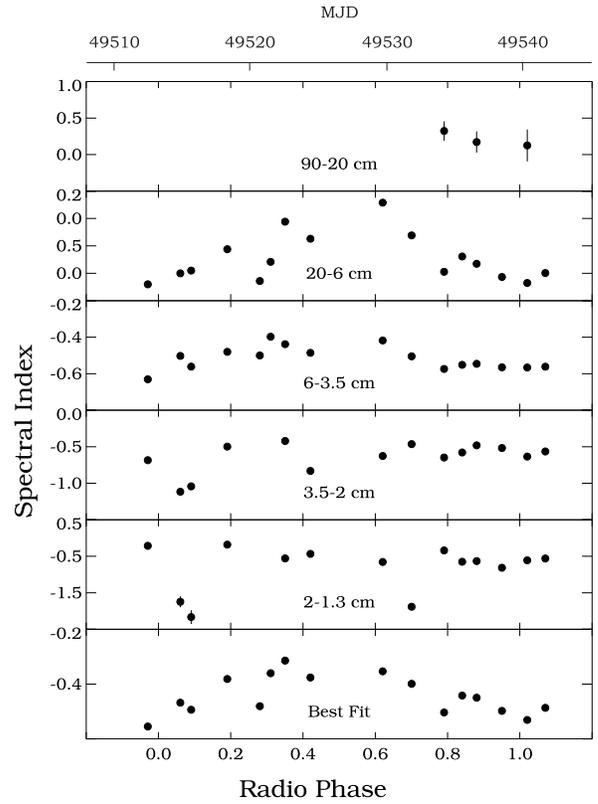}
\caption{\label{fig:specind}
Time evolution of the spectral indices of LSI +61$^\circ$303 during
the outburst of 1994 June-July. They have been computed
between all adjacent wavelength pairs.}
\end{figure}

To test the super-Eddington accretion/particle-injection model we assume the
radio emission can be attributed to synchrotron radiation from relativistic
electrons in hot ionized clouds (plasmons), with the electrons
accelerated either by shocks in gas dynamical expansions or by
the magnetosphere of
the compact companion (Penninx \markcite{penni89}1989).   The radio light
curves of LSI~+61$^{\circ}$303 shown
in Figures~\ref{fig:ossegbilc} and \ref{fig:curves} have a shape dominated by
the occurrence of the expected periodic radio outburst, between radio phases
0.4 and 1.0.  This main flaring event was preceded by at least two minor
outbursts, at phases 0.1 and 0.3.  In addition, the 1.3 cm data suggests that
there was a third minor outburst, centered around phase 0.8, whose radio
emission overlaps with the decay of the main event. The expected
radio spectrum
is a non-thermal power law with a negative spectral index $\alpha$
($S_{\nu}\propto \nu^{\alpha}$).  In general terms, the
spectral indices shown in Figure~\ref{fig:specind} do agree with
this expectation.  However,
deviations from simple power law behavior are evident in our data (see 
Figure~\ref{fig:specevol}), and they seem to be clearly related with the state
of radio activity of the source.  In this way, during the observed periods of
low radio activity (phases 0.0-0.3 and 0.9-0.1), the radio spectrum between
20, 6 and 3.5 cm is well represented as a simple power law of spectral index
$\alpha\simeq-0.5$.  However, during the flaring interval (phases 0.3-0.9),
the spectral indices undergo noticeable variations especially at both ends of
the spectrum.

The changes of $\alpha_{20-6 {\rm cm}}$ are probably due to opacity
effects, i.e., the development of an optically thick synchrotron
spectrum during the rise of the main outburst.  This kind of behavior
has been previously observed in other outbursts of this source (Taylor
\& Gregory, \markcite{taylo84}1984; Taylor et al., \markcite{taylo95}1995)
and may be attributed to
synchrotron self-absorption in the radio emitting plasmon during the
early stages of the outburst.  The plots of spectral index in
Figure~\ref{fig:ossegbilc}
and $\alpha_{20-6 {\rm cm}}$ in
Figure~\ref{fig:specind} suggest that the maximum value of $\alpha$
occured
between phase 0.5--0.6, just in the middle of the flux density rise.
On the other hand, in spite of the scarcity of the data at 90 cm, the
$\alpha_{90-20 {\rm cm}}$ plot has a variability trend similar to that
of $\alpha_{20-6 {\rm cm}}$ and is the only spectral index observed to
be positive even during the decay of the main outburst.  This suggests
that the turnover frequency of the 
LSI~+61$^{\circ}$303 synchrotron spectrum is likely
to occur between 0.3 and 1.4 GHz throughout all the radio period.

At intermediate frequencies, $\alpha_{6-3.5 {\rm cm}}$ remains quite
constant and negative at the $-0.5$ value, indicating that 
LSI~+61$^{\circ}$303 is
always optically thin between 6 and 3.5 cm, even during the onset of the
radio outburst.  As suggested by Taylor \& Gregory \markcite{taylo84}(1984) and modelled by
Paredes et al.  \markcite{pared91}(1991), this behavior is consistent with continuous
injection of relativistic electrons in order to account for the
flux density rise at optically thin frequencies.

At the highest frequencies, both $\alpha_{3.5-2 {\rm cm}}$ and
$\alpha_{2-1.3 {\rm cm}}$ are always negative but not constant.  The
spectral index changes at these optically thin frequencies are likely to
be related to the energy evolution of the relativistic electrons
undergoing energy losses of different type.  The energy loss mechanisms 
that have the strongest effect on the
high frequency steepening of the radio spectrum are synchrotron radiation
and inverse Compton losses (Pacholczyk, \markcite{pacho70}1970), due to their energy square
dependence.  Since an electron with
energy $E$ radiates mainly at a characteristic frequency $\nu_c \propto
E^2$, the effect of energy losses will be noticeable first at higher
frequencies, where the radiating electrons will lose their energy more
rapidly.  In this way, once the injection process is over and energetic
electrons are no longer replaced, one expects a progressive high
frequency steepening of the spectrum during the decay of the outburst
(Paredes, Peracaula \& Mart\'{\i} \markcite{pared97b}1997).

This expectation is actually confirmed by observations, especially by the
$\alpha_{2-1.3 {\rm cm}}$ plot of Figure~\ref{fig:specind}, where the deep
minima at phases 0.1 and 0.7 correspond to significant steepening of the
spectrum during the decay of both the first minor and main outburst.
There is also evidence of steepening during the decay of the second and
third minor outbursts, at phases 0.35 and 0.95, but this is not so
pronounced. In addition, the $\alpha_{3.5-2.0 {\rm cm}}$ plot
has also indications of similar steepening, but the dips of this curve
are not so clear and they seem to occur slightly later in time
than those of $\alpha_{2-1.3 {\rm cm}}$, as expected as we go towards
lower frequencies.

It is also important to remark that, following each $\alpha_{2-1.3 {\rm
cm}}$ steepening, the recovery of this spectral index leaves the 1.3 cm
flux density at a value higher than expected from a simple extrapolation
using the nearby $\alpha_{3.5-2.0 {\rm cm}}$.  This can be seen several
times in the Figure~\ref{fig:specevol} plots, at phases 0.97, 0.19, 0.42 and
0.79.  We note that all these phases correspond to the onset of a
flaring event or to its very early rise in the case of the main
outburst.  So, we speculate that these 1.3 cm flux density ``excesses''
could be attributed to the first stages of the particle injection
process, when very fresh and short-lived highly energetic electrons are
being produced and injected into the plasmon.  

Overall, particle-injection from super-Eddington accretion
can apparently explain the shape of the radio light curve observed in 
the system. However, the paucity of X-ray emission
from LSI~+61$^\circ$303  both in the soft (Taylor et~al. 1996)
and hard bands (Tavani et~al. 1996) does not agree with the
expected flux from a compact object accreting at super-Eddington
rates. Recently discovered superluminal jet X-ray transients 
GRS 1915+105 (Mirabel \& Rodriguez 1994)
and GRO~J1655-40
(Hjellming \& Rupen 1995) are characterized by strong hard
X-ray outbursts in coincidence with relativistic plasmoid  ejections.
BATSE continuous monitoring of 2CG~135+1 in the hard X-ray range
has established an upper limit of $\sim 10$mCrab (20-100~keV)
during the period April 1991-January 1995 (Tavani et al. 1996).
This BATSE upper limit is more than two orders of magnitude lower
than the expected flux from a super-Eddington accreting source 
similar to the Galactic superluminal X-ray transients (at the distance
of LSI~+61$^\circ$303).
Furthermore, the observed small column density of LSI~+61$^\circ$303,
$N_H \sim 6\cdot 10^{21} \rm \, cm^{-2}$
as deduced from ROSAT (Goldoni \& Mereghetti 1995, 
Taylor et al., 1996) and ASCA observations (Leahy et~al. 
\markcite{leahy96}1996),
is consistent with negligible intrinsic absorption at the source.
No  Fe X-ray line emission was detected  by ASCA in
coincidence with a radio flare (Leahy et al. \markcite{leahy96}1996).
We therefore conclude that
if 2CG~135+1 is related to an accreting source, its properties
must be quite different  from those of GRS~1915+105 and GRO~J1655-40
or other currently known accreting  X-ray  sources.

\subsection{Optical and Infrared emission}

Paredes et al. \markcite{pared94}(1994) proposed that various features in the optical
spectra of LSI +61$^\circ$303 and its infrared flux were modulated 
with the radio period of the system.  Our optical and infrared
observations allow us to begin to test those claims and search for 
evidence of mass ejections from the system correlated with the radio
outbursts, as would be predicted by the super-Eddington model.

\subsubsection{H$\alpha$ measurements}

\begin{figure}
\plotone{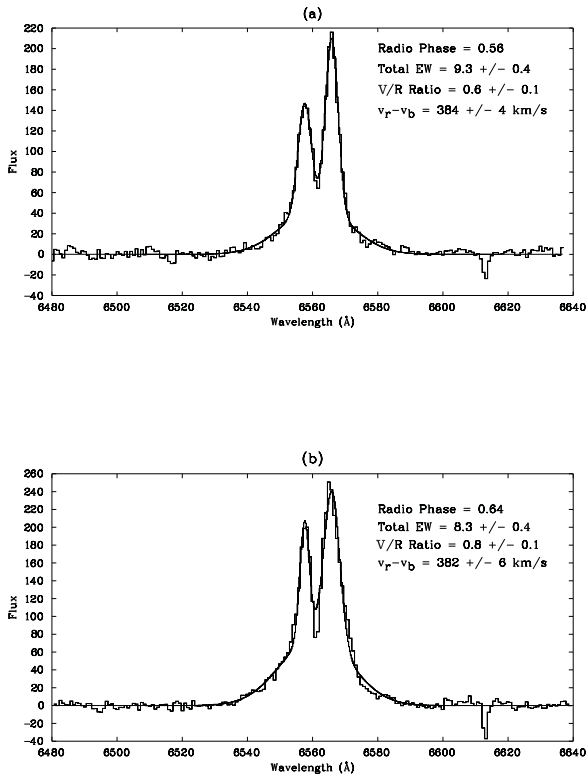}
\caption{\label{fig:halphaspec}
Details of the H$\alpha$ profiles taken at two different
binary phases.  The vertical scales are in arbitrary flux units.}
\end{figure}

Figure \ref{fig:halphaspec} shows the H$\alpha$ line profiles obtained during the
observations.  The profile is double peaked, with the
red peak appreciably stronger than the blue in both spectra.  In
addition, broad wings to the H$\alpha$ line are apparent in both
spectra.  A general discussion of the appearance of the spectrum of
this source, with particular emphasis on the H$\alpha$ line, is given
by Gregory et al. \markcite{grego79}(1979) and Paredes et al. 
\markcite{pared94}(1994).  
Comparison of our spectra with those presented by Paredes et
al. \markcite{pared94}(1994), which cover all radio phases, shows the basic appearance
unchanged since 1989.  
A more quantitive comparison of the new spectra, both with
each other and with the historical spectra, may be made by fitting
appropriate functions to the profile.  However, in carrying out this
comparison it is important to remember that many isolated
(i.e. non-binary) Be stars show variations very similar to those
described here, and that any attempt to correlate these with the
orbital phase using the small sample of data presented here will
necessarily be unreliable.  This is, however, the first optical data
obtained simultaneously with high energy observations, and is, 
as such, of interest.

The H$\alpha$ fit  was carried out using the
DIPSO emission line fitting (ELF) routines described by Howarth \& Murray
\markcite{howar91}(1991).  We found that the H$\alpha$ line is
well represented by a
combination of red and blue narrow (FWHM$\sim5$ \AA) and single
broad (FWHM$\sim20 $\AA) Gaussian components.  Based on similar fits to their
data, Paredes et al. \markcite{pared94}(1994) discuss the
possible origin of the three components in a disk surrounding the Be
star, with the two narrow components arising naturally from rotation
of the disk and the broad component due to electron scattered photons. 
The derived fit parameters
are listed in Table~\ref{halphatab}, and the fits themselves overplotted on the
spectra of Figure~\ref{fig:halphaspec}.  Note that no attempt was made when
calibrating the spectra to account for `slit' and other losses.  The absolute
calibration of the flux values from night to night will therefore vary
considerably.  To allow for this we list the interpolated 
continuum flux value at the centre of the H$\alpha$ line for each
spectrum.  Using these values allows the conversion of the
measured fluxes into 
equivalent widths (which are independent of the flux calibration).

From the table no significant changes are found 
in the central wavelengths of the
three Gaussians between the observations.  However the FWHM of the red
component increases from 4.5 to 5.8 {\AA} between radio phases 0.56 and 0.64.
Similar behavior was seen by Paredes et al. \markcite{pared94}(1994), who observed an
increase in FWHM$_r$ from 6 to 8 {\AA} between phases 0.5 to 0.7.
They suggested that this broadening of the red peak
near radio maximum may be caused by an unresolved component of gas
ejected from the system at that time.    
We note however that no
significant change in the separation of the red and blue peaks
(expressed on the figure in terms of the velocity $v_r$-$v_b$)
is seen in our data, which might be expected if this were the case.

The V/R ratio is a standard quantity used in the analysis of Be star
spectra, and is defined as the ratio of the peak
fluxes of the blue and red components.  In this case we use the peaks
of the fitted components, and find the ratio increases from 
$0.6\pm0.1$ to $0.8\pm0.1$ between
radio phases 0.56 to 0.64.   This result is consistent with a constant V/R
value.

The total H$\alpha$ equivalent width
derived from the fit parameters is also shown on
Figure~\ref{fig:halphaspec}.  A slight decrease in the H$\alpha$ EW by $\sim1$ {\AA} 
between the two
spectra is apparent.  This is again consistent with the pattern
for similar radio
phases reported by Paredes et al. \markcite{pared94}(1994), and will be further
discussed below in conjunction with the H$\beta$ and H$\gamma$
observations of the source.

\subsubsection{H$\beta$ and H$\gamma$ measurements}

In Figure \ref{fig:hbetaspec} 
we plot our observations of the H$\beta$ and H$\gamma$
lines from the source.  H$\beta$ shows a typical Be `shell' spectrum
with a deep central absorption core and two prominent wings.  In both H$\beta$
spectra the red wing is considerably stronger than the blue.  We also
note possible evidence for a weak, very broad blue wing,
presumably analogous to the broad component in
the H$\alpha$ spectrum, which is also slightly blue shifted according
to its derived central wavelength.  The
H$\gamma$ line is in absorption, although it is likely that some 
`in-filling' of the line due to emission is present.

Due to
the poorer signal-to-noise ratio of these spectra, it is not feasible
to fit Gaussian profiles to either the H$\beta$ or H$\gamma$
lines.  Instead we measure total line equivalent widths using the
FIGARO routine ABLINE
(Robertson \markcite{rober86}1986).  The resulting EWs are given on the plots of the spectra.
A similar change is apparent in the EWs of both the H$\beta$ and
H$\gamma$ lines between the two nights, namely a decrease in the (emission)
equivalent width of both lines by $\sim0.5$ {\AA}.  This occurs
between radio phases 0.60
and 0.67.  In the previous section we noted that the H$\alpha$ EW
decreased by $\sim1$ {\AA} between phases 0.56 and 0.64.  It therefore
appears that the emission strengths of all of the Balmer
recombination lines decreased steadily from radio phase 0.56 to 0.67.
Whether this behavior is in fact a function of orbital phase, or
merely represents uncorrelated variations in the circumstellar
conditions, is naturally uncertain with the present dataset.  However,
it is encouraging that changes can be reliably measured on such a short
time scale, and shows that a spectroscopic study of the system
throughout a complete radio cycle would be extremely valuable.  Such a study
would also allow a redetermination of the orbital parameters of the
system, which are currently based on the determination of Hutchings \&
Crampton \markcite{hutch81}(1981).

\subsubsection{Infrared photometry}

As shown in Figure~\ref{fig:irlc}, the source brightens in IR between radio phases 
0.44 and 0.67.  This may be due to either 
some form of re-processing of the radio flux
during the outburst (a negative search for similar behavior in the
source EXO2030+375 is reported by Norton et al. \markcite{norto94}1994), or some
type of eclipse within the system.  
Examination of the folded infrared
light-curve presented by Paredes et al. \markcite{pared94}(1994) shows that the infrared
variability of the source is more likely to be the result of some form
of eclipse in the system, with the eclipse occurring between radio
phases $\sim 0.3$ and $\sim 0.5$.  The new data presented here do not
allow us to either confirm or deny the eclipse interpretation.
A complete set of
observations through a single phase cycle will be necessary to 
properly understand the infrared variability of the system.  In
particular, it is important to note that superimposed on any 
possible orbital
modulation will be long term variations in the infrared flux from the
system.  Therefore any inferences drawn from the folded light
curve of Paredes et al. \markcite{pared94}(1994) must be treated with some caution.

%
%
%
%
%
%
%
%
%
%
%
%
%
%
\subsubsection{Optical line emission and high energy flux}

One possible mechanism to account for the H$\alpha$ variability we observe is
reprocessing of the X-ray flux from the system. Apparao 
(\markcite{appar91}1991) has shown that the presence of an X-ray emitting
source embedded in a Be star wind should create an H {\sc ii} region in the
wind which would emit optical line and continuum radiation. Using our total
H$\alpha$ flux density of $\sim 3700$mJy,  the radio derived distance to the
system (2.0 kpc - Frail \& Hjellming \markcite{frail91}1991)  and an
extinction derived from the color excess  ($E(B-V)\sim1.13$ Paredes \&
Figueras \markcite{pared86}1986)  we derive a total H$\alpha$ emission from
the system of $\sim10^{35}$ ergs/s. The ASCA 2 to 10 keV flux is however
substantially less at 5$\times10^{33}$ ergs/s (Leahy et al.
\markcite{leahy96}1996).  Therefore a maximum H$\alpha$ variability of only 5
per-cent of the line EW is likely to be attributable to reprocessing of high
energy flux.  In addition, we note the efficiency for the processing of X-ray
to H$\alpha$ flux presented by Apparao (\markcite{appar91}1991), which (although obviously
dependant on the physical conditions and the input X-ray spectrum)  is
generally $\sim 0.05$. A more reasonable estimate of the expected effect of
reprocessing in this system may therefore be $\sim 0.2$ percent of the EW. 

We observe a total decline in H$\alpha$ EW of around 10 per-cent over 2 days
in the system and the evidence from the H$\beta$ and $\gamma$ lines is that
the Balmer series emission declined further the following day.  We therefore
estimate a total decline of 15 per-cent in the H$\alpha$ EW over three days,
substantially higher than our derived maximum of 5 percent from the previous
paragraph. We therefore conclude that the H$\alpha$ variability we observe
from the system is unlikely to be due to reprocessing of the variable X-ray
flux, and the most likely explanation is that it is simply the normal
variability associated with the Be phenomenon.  An analogous calculation made
for the IR flux from the system yields the same result.
%
%
%
%
%
%

\section{Conclusions}

This paper has discussed the results of a multiwavelength monitoring campaign
on the radio/X-ray/Be binary LSI +61$^\circ$303 and the $\gamma$-ray source
2CG 135+1.  The original aim of the campaign was 
to prove the
association between the two sources, either by spectral or temporal
signatures.  We observed the source three times using the OSSE detector on
CGRO, but only detected it during the first observation.
The OSSE data for the field containing 2CG~135+1 have
insufficient signal-to-noise ratio to search for any correlation  with the
radio light curve of LSI +61$^\circ$303.
 A  comparison of the OSSE spectrum
from the direction of  2CG~135+1 with 
spectra of LSI +61$^\circ$303 at  lower and higher energies 
(as determined by ROSAT, ASCA and  EGRET)
does not allow an unambiguous association between these two sources.
We notice that a  substantial
part of the high-energy flux detected by  OSSE and COMPTEL 
might originate  from QSO 0241+622.  Future observations of the
2CG 135+1 field
with OSSE (using different collimator orientations) and with the ROSSI X-ray
Timing Explorer and SAX  (which have a smaller fields of view) 
may help resolve this issue.


We also discussed the
implications of our 
radio, optical and infrared 
observations of LSI +61$^\circ$303.
We have shown
that the models proposed to explain the features of the
LSI +61$^\circ$303 system
(super-Eddington neutron star accretion and  a non-accreting 
shock-driven young pulsar system) 
are able to
qualitatively explain the features of the 
observed multifrequency radio lightcurves.
More data are required to resolve the outstanding issues concerning
this system.
We also notice that
the study of the  radio spectrum of LSI +61$^\circ$303  deserves
 more
attention in order to better understand the physics of the acceleration
of radio emitting electrons.  In particular, it would be very
interesting to carry out  multi-frequency monitoring of a full outburst
including simultaneously both the the centimeter and millimeter wavelength
domain.  The optical spectroscopy and infrared photometry
have shown that genuine night-to-night variability is also present.  
An orbit-long monitoring campaign at these wavelengths 
would be very useful
in determining whether the observed variability was truly correlated
with orbital phase.

\section{Acknowledgements}

Research  partially supported by the GRO Guest Investigator Program
(grant NAG-5-2729).  JMP acknowledges support by DGICYT (PB94-0904). JM is
supported by a postdoctoral fellowship of the Spanish Ministerio de
Educaci\'on y Cultura. The NRAO is operated by Associated Universities, Inc.,
under cooperative agreement with the National Science Foundation.  The JKT is
operated by the Royal Observatory, Greenwich on behalf of the UK PPARC.  The
TCS is operated by the Instituto de Astrof\'{\i}sica de Canarias, Tenerife.
Some of the data reduction for this paper was carried out using the
Southampton University STARLINK node.


\onecolumn
\clearpage
\begin{table}
\caption{\label{tbl:osselog} OSSE Observations of 2CG 135+1}
\vspace{10pt}
\begin{tabular}{llll}
VP & Dates & Radio Phase & Exposure \\
\tableline
325 & April 26 -- May 10 1994 & 0.31 -- 0.84 & 3.30$\times10^5$ s\\
330 & June 10 -- June 14 1994 & 0.01 -- 0.16 & 0.98$\times10^5$ s\\
332 & June 18 -- July 5 1994 & 0.32 -- 0.95 & 3.60$\times10^5$ s\\
\end{tabular}
\end{table}

\begin{table}
\caption{\label{tbl:osseflux} OSSE Flux measurements and upper limits}
\vspace{10pt}
\begin{tabular}{llllll}
 & \multicolumn{5}{c}{Flux (photons-cm$^{-2}$-s$^{-1}$-MeV$^{-1}$)}\\
 VP & 0.05--0.1 MeV & 0.1--0.17 MeV & 0.17--0.3 MeV & 0.3--1.5 MeV & 1.5--10 MeV\\
\tableline
325 & $4\pm2\times10^{-3}$ & $1.2\pm0.5\times10^{-3}$ 
& $1.0\pm0.5\times10^{-3}$ & $1.4\pm1.5\times10^{-4}$
& $0.5\pm1.6\times10^{-5}$\\
330 & $2\pm3\times10^{-3}$ & $-0.2\pm0.9\times10^{-3}$ 
& $0.2\pm0.7\times10^{-3}$ & $1\pm2\times10^{-4}$
& $-0.3\pm2.5\times10^{-5}$\\
332 & $3\pm2\times10^{-3}$ & $0.6\pm0.5\times10^{-3}$ 
& $-0.5\pm0.4\times10^{-3}$ & $0.8\pm1.3\times10^{-4}$
& $1\pm1\times10^{-5}$\\
\end{tabular}
\end{table}

\begin{table}
\caption{\label{radiodata} Multifrequency VLA Radio Observations of 
LSI +61$^\circ$303}
\vspace{10pt}
\begin{tabular}{crrrrrrr}
             &          &                 &                  &                 &                   &                &                  \\
 MJD  &   Radio  & $S_{\rm 90 cm}$ & $S_{\rm 20 cm}$  & $S_{\rm 6 cm}$  & $S_{\rm 3.5 cm}$  &$S_{\rm 2.0 cm}$& $S_{\rm 1.3 cm}$ \\
   &   Phase  &      (mJy)      &      (mJy)       &      (mJy)      &      (mJy)        &     (mJy)      &     (mJy)        \\
             &          &                &                  &                 &                   &                &                  \\
\tableline
             &          &                 &                  &                 &                   &                &                  \\
   49512.93   &   0.97   &                 &  $ 54.4 \pm 0.4$ & $ 30.5 \pm 0.1$ & $ 21.71 \pm 0.05$ & $14.8 \pm 0.2$ & $13.5 \pm 0.5$   \\
   49515.34   &   0.06   &                 &  $ 69.8 \pm 0.5$ & $ 43.1 \pm 0.1$ & $ 32.87 \pm 0.07$ & $17.6 \pm 0.4$ & $ 8.3 \pm 0.5$   \\
   49516.34   &   0.09   &                 &  $ 48.7 \pm 0.4$ & $ 30.8 \pm 0.1$ & $ 22.76 \pm 0.05$ & $12.7 \pm 0.2$ & $ 5.0 \pm 0.4$   \\
   49518.96   &   0.19   &                 &  $ 35.9 \pm 0.4$ & $ 27.4 \pm 0.1$ & $ 21.15 \pm 0.05$ & $16.0 \pm 0.2$ & $14.8 \pm 0.4$   \\
   49521.34   &   0.28   &                 &  $ 57.7 \pm 0.5$ & $ 33.3 \pm 0.1$ & $ 25.43 \pm 0.05$ &                &                  \\
   49522.17   &   0.31   &                 &  $ 65.3 \pm 0.4$ & $ 44.6 \pm 0.1$ & $ 36.00 \pm 0.05$ &                &                  \\
   49523.16   &   0.35   &                 &  $ 43.2 \pm 0.4$ & $ 42.0 \pm 0.1$ & $ 33.16 \pm 0.06$ & $26.2 \pm 0.2$ & $20.6 \pm 0.5$   \\
   49525.05   &   0.42   &                 &  $ 75.4 \pm 0.4$ & $ 63.1 \pm 0.1$ & $ 48.56 \pm 0.08$ & $30.5 \pm 0.2$ & $25.3 \pm 0.7$   \\
   49530.28   &   0.62   &                 &  $146.6 \pm 0.5$ & $168.5 \pm 0.1$ & $134.50 \pm 0.07$ & $94.7 \pm 0.2$ & $71.4 \pm 0.4$   \\
   49532.28   &   0.70   &                 &  $158.6 \pm 0.5$ & $136.7 \pm 0.1$ & $104.13 \pm 0.05$ & $80.3 \pm 0.3$ & $35.7 \pm 0.6$   \\
   49534.87   &   0.79   & 106$\pm$21      &  $172.4 \pm 0.5$ & $107.9 \pm 0.1$ & $ 79.19 \pm 0.06$ & $55.1 \pm 0.2$ & $47.6 \pm 0.5$   \\
   49536.06   &   0.84   &                 &  $136.2 \pm 0.4$ & $ 97.5 \pm 0.1$ & $ 72.43 \pm 0.05$ & $52.4 \pm 0.2$ & $39.6 \pm 0.4$   \\
   49537.12   &   0.88   & 106$\pm$23      &  $137.3 \pm 0.4$ & $ 92.2 \pm 0.1$ & $ 68.70 \pm 0.05$ & $52.5 \pm 0.2$ & $40.0 \pm 0.4$   \\
   49539.13   &   0.95   &                 &  $115.7 \pm 0.4$ & $ 69.2 \pm 0.1$ & $ 51.03 \pm 0.05$ & $38.2 \pm 0.2$ & $26.9 \pm 0.5$   \\
   49540.96   &   0.02   &  61$\pm$20      &  $ 73.7 \pm 0.5$ & $ 41.8 \pm 0.1$ & $ 30.81 \pm 0.05$ & $21.6 \pm 0.2$ & $16.6 \pm 0.4$   \\
   49542.11   &   0.07   &                 &  $ 58.8 \pm 0.4$ & $ 36.4 \pm 0.1$ & $ 26.89 \pm 0.05$ & $19.6 \pm 0.2$ & $15.4 \pm 0.4$   \\
             &          &                 &                  &                 &                   &                &                  \\
\end{tabular}
\end{table}

\begin{table}
\caption{\label{optlog}Log of Optical Spectroscopy}
\vspace{10pt}
\begin{tabular}{llll}
MJD & Radio Phase & $\lambda$ (\AA) & T$_{\rm exp}$ (s) \\
\tableline
49528.7 & 0.56 & 6030 -- 7070 & 1200 \\
49529.7 & 0.60 & 4200 -- 5200 & 1800 \\
49530.7 & 0.64 & 6070 -- 7040 & 1500 \\
49531.7 & 0.67 & 4200 -- 5200 & 1800 \\
\end{tabular}
\end{table}

\begin{table}
\caption{\label{irtable}Infrared Photometry}
\vspace{10pt}
\begin{tabular}{lllll}
MJD & Radio Phase & J & H & K \\
\tableline
49525.5 & 0.44 & 8.68$\pm$ 0.02 & 8.28$\pm$ 0.02 & 7.97$\pm$ 0.02 \\
49526.5 & 0.48 & 8.62$\pm$ 0.02 & 8.21$\pm$ 0.02 & 7.93$\pm$ 0.02 \\
49527.5 & 0.52 & 8.58$\pm$ 0.03 & 8.19$\pm$ 0.02 & 7.89$\pm$ 0.03 \\
49530.5 & 0.63 & 8.60$\pm$ 0.05 & 8.19$\pm$ 0.04 & 7.86$\pm$ 0.03 \\
49531.5 & 0.67 & 8.62$\pm$ 0.04 & 8.22$\pm$ 0.02 & 7.89$\pm$ 0.02 \\
49535.5 & 0.82 & 8.63$\pm$ 0.12 & 8.23$\pm$ 0.12 & 7.94$\pm$ 0.08 \\
\end{tabular}
\end{table}
\newpage

\begin{table}
\caption{\label{halphatab}H$\alpha$ fit parameters}
\vspace{10pt}
\begin{tabular}{lll}
 MJD & 49528.7 & 49530.7 \\ 
\tableline
 $\lambda_b$ (\AA)&   $6657.53\pm0.07$ &$6657.56\pm0.09$ \\
 $\lambda_r$ (\AA)& $6565.94\pm0.04$ & $6565.93\pm0.09$\\
 $\lambda_c$ (\AA)& $6560.88\pm0.40$ & $6561.12\pm0.49$\\
 FWHM$_b$ (\AA)& $4.5\pm0.2$ &$3.5\pm0.2$ \\
 FWHM$_r$ (\AA)& $4.5\pm0.2$ & $5.8\pm0.3$ \\
 FWHM$_c$ (\AA)& $20.6\pm1.1$ & $26.7\pm1.4$ \\
 Peak$_b$ (mJy) &$98\pm5$ &$135\pm8$\\
 Peak$_r$ (mJy) &$165\pm5$&$172\pm7$\\
 Peak$_c$ (mJy) &$47\pm$5 & $78\pm7$\\
 Flux$_b$ (mJy) &$474\pm42$ & $502\pm49$ \\
 Flux$_r$ (mJy) &$794\pm41$ & $1060\pm80$ \\
 Flux$_c$ (mJy) &$1050\pm85$ & $2180\pm118$ \\
 Continuum (mJy)&$250\pm10$   & $450\pm10$ \\
\end{tabular}  
\end{table}

\twocolumn

\clearpage

\newpage
\newpage
\newpage
\newpage
\newpage
\newpage
\newpage
\newpage
\end{document}